\newif\ifrevtex
\newif\ifsubmission
\newif\iftwocolumn

\revtextrue   
\revtexfalse  

\submissiontrue 

\ifrevtex\else\submissionfalse\twocolumntrue\fi

\ifrevtex
  \ifsubmission
    \documentstyle[floats,prl,aps,preprint,times]{revtex}
    \twocolumnfalse
  \else
   \documentstyle[floats,prl,aps,twocolumn,times]{revtex}
   \tighten
  \fi

  \newcommand{\affiliation}[1]{\address {#1}}
  \renewcommand\ensuremath\relax
  \newcommand\eqref[1]{(\ref{#1})}

  \newcommand\citep[1]{\cite{#1}}
  \renewcommand{\usepackage}{\RequirePackage}
  \usepackage{times,graphicx,psfrag,subfigure,url}
\else
  \documentclass[pre,aps,twocolumn,tightenlines]{revtex4}
  \usepackage{times,psfrag,amsmath,subfigure}
  \usepackage[draft]{graphicx}
  \DeclareMathAlphabet{\mathrmb}{OT1}{ptm}{b}{n}
  \DeclareMathAlphabet{\mathsfb}{OT1}{phv}{b}{n}
\fi

\ifsubmission
  \newcommand{\gpfig}[1]{\relax}
  \renewcommand{\includegraphics}[1]{\relax}
  \renewcommand{\caption}[1]{\relax}
  \newcommand{\xlabel}[1]{\relax}
  \newcommand{\ylabel}[1]{\relax}
  \usepackage[nolists,nomarkers]{endfloat} 
  
\else
  \graphicspath{{eps/}}
  \newcommand{\xlabel}[1]{\psfrag{xlabel}[][]{#1}}
  \newcommand{\ylabel}[1]{\psfrag{ylabel}[b][t]{#1}}
  \iftwocolumn
    \newcommand{\gpfig}[1]{%
      \includegraphics[width=8.6cm]{#1}
      }
  \else
    \usepackage[nolists,nomarkers]{endfloat} 
    \newcommand{\gpfig}[1]{%
      \thispagestyle{empty}
      Sunthar and Kumaran (Spontaneous phase separation \ldots)
      \vspace{3cm}\par
      \includegraphics[width=8.6cm]{#1}
      \\[\baselineskip]\vspace{3cm}
      }
   \fi
\fi

\usepackage{xspace}

\newcommand{\Exp}[1]{{\rm e}^{#1}}
\newcommand{\textfrac}[2]{\ensuremath{#1/#2}}

\newcommand{\Figref}[1]{Fig.~\ref{#1}}

\newcommand{\Eqref}[1]{Eq.~\eqref{#1}}

\newcommand{\granuhome}{%
  \urlstyle{rm}
  \url{http://chemeng.iisc.ernet.in/research/r-granular.html}}

\newcommand\param[3]{%
  The parameters used for this simulation are: $\ndp=#1$,
  $\epsilon=#2$, and $\dens = #3$}
\newcommand{\ndp}{\ensuremath{N \sigma}\xspace}
\newcommand{\dens}{\ensuremath{\nu^{\circ}}\xspace}

\newcommand{\waves}{waves}
\newcommand{\nongauss}{nongauss}
\newcommand{\phasessim}{phases-sim}
\newcommand{\phasesexpt}{phases-expt}

\begin{document}

\title{Spontaneous Phase Separation in a Vibro-fluidized Granular
  Layer} 

\author{P Sunthar and V Kumaran}
\affiliation{Department of Chemical Engineering, \\ 
  Indian Institute of Science, Bangalore, 560 012, India.}

\ifrevtex
  \maketitle  
\fi

\begin{abstract}
  A new type of two phase coexistence in a vibrated granular material
  is reported in the limit where energy transfer between the particles
  and the bottom wall occurs due to discrete collisions.  A
  horizontally homogeneous bed develops inhomogeneities in which a
  dense and a dilute phase separate and coexist. The dilute region
  resembles a gas, while the dense region is like a liquid and a velocity
  distribution very different from a Gaussian.
\end{abstract}

\pacs{45.70.Qj, 05.65.+b, 45.70.-n}

\ifrevtex
\else
  \maketitle  
\fi
Granular materials occur in diverse sizes and shapes.  This leads to a
variety of complex patterns in collective motions of grains, which are
of relevance to predicting natural phenomena as well as for process
design in industry \citep{jae-nag96}.  Though the interactions between
grains are complex in nature, many complex patterns are reproduced by
simple microscopic interactions, such as hard sphere inelastic
collisions.  Of recent interest are granular materials subjected to
vertical vibrations.  When the amplitude of bottom wall vibrations is
comparable to the bed height, the top surface becomes unstable and
forms a periodic wavy pattern with peaks and troughs \citep{\waves}.
An unusual type of excitation is the ``oscillons'' reported in
\citep{umbanetal96} where the excitation is localized.  In general it
is argued that \citep{meloetal95} the various patterns observed in a
vibrated granular layers are a result of interaction between standing
waves.  Several of these patterns---squares, stripes, hexagons, kinks,
etc.---have been recovered have by a simple model of interacting waves
\cite{venkat-ott98}.

Eggers \cite{eggers99} has simulated a vibrated bed under gravity with
two compartments separated by a wall which allows particles from one
side to the other through a small slit at a given height. For vigorous
shaking, the populations in the two halves remain equal, but at a
lesser rate of shaking, the bed separates into stable coexistence of a
high and a low density phase.  It is also shown that such a stationary
state can be captured by a flux balance of particles from one side to
the other calculated by the simple kinetic theory of
\cite{kum98:vibscal}.  Occurrence of large scale inhomogeneities in
industrial vibrated beds is reported in \cite{buevich-ryzhkov80},
which seem to be similar in nature to a two phase coexistence, but the
instability disappears when the surrounding air medium is evacuated.
Phase separations have been observed in two dimensional horizontal
plane simulations \citep{\phasessim} and experiments
\citep{\phasesexpt}.  The particles are restricted to move only in the
horizontal two dimensional plane, and the coexistence is between two
phases whose static structure are an ordered (collapsed) phase and a
liquid-like phase (which is referred to as ``gas-like'' in
\cite{\phasesexpt}).  For a mono-layer of particles, the the ratio
$\Gamma$ of the maximum acceleration of the wall and the acceleration
due to gravity in \cite{\phasesexpt} is less than unity.
  
A new type of phase coexistence phenomenon observed in two dimensional
computer simulations of a vertically vibrated granular material is
reported here.  The parameter regime where this is observed differs
from that of previous studies in two respects.
\begin{enumerate}
\item These are observed in a low density regime, where the area
  fraction of the particles is small, i.e., the mean free path is
  large compared to the particle diameter.  This is in contrast to
  previous studies \citep{\phasesexpt}
  where the mean free path less than the particle diameter.
\item In the present case, the frequency of the base vibration is
  large compared to the frequency of oscillation of the bed, and
  amplitude is small compared to the mean free path of the particles.
  Consequently, the amplitude and frequency of the base are not
  relevant parameters in determining the state of the bed as in
  \citep{\waves}, and the only relevant parameter is the energy
  transfer from the base to the particles in discrete particle
  collisions.
\end{enumerate}

Since we are interested in the dilute limit, it is useful to
characterize the system by the parameters relevant for a dilute
vibro-fluidized bed \citep{kum98:vibscal} in the limit where the area
fraction is small and the coefficient of restitution is close to $1$.
The relevant parameters are $\ndp$, which gives the number of
mono-layers of particles at rest, and the parameter $\epsilon =
(U_{0}^{2} / T)$, which is a measure of the ratio of the dissipation
over time scales comparable to the time between collisions and the
average energy of a particle. Here $N$ is the number of particles per
unit width, $\sigma$ is the particle diameter, $U_{0}^{2}$ is the mean
square velocity of the vibrating surface and $T$, the `granular temperature',
is the mean square velocity of the particles (the mass of a particle is set
equal to $1$).  It was shown that in the limit $(1 -
e^{2}) \ll 1$, the parameter $\epsilon$ is given by
\begin{equation}
  \label{eq:T0}
\epsilon \equiv \frac{U_{0}^{2}}{T} = \frac{\pi \, \ndp \, (1 -
  e^{2})}{2 \sqrt{2}}.
\end{equation}
The area fraction of particles $\nu$ in the dilute limit is
exponentially decaying in the vertical direction, and is given by
$ \nu = \dens \, \Exp{ - \textfrac{g z}{T}}, $ 
where $\dens = \pi \, N \, g \, \sigma^{2} / (4 T)$ is the area
fraction of particles at the bottom of the bed.  The area fraction at
the bottom is kept at a constant value $\dens = 0.077$ in the simulations.

The other parameters usually used for characterizing pattern formation
in vibrated granular systems are not relevant for the present simulations.
\begin{enumerate}
\item The ratio of characteristic frequency of vertical oscillation of the bed
  $(g / \sqrt{T})$, and frequency $\omega_{w}$ of the bottom wall
  oscillations, is maintained at a constant value of $0.01$ in the
  simulations, indicating that there are no correlated collisions of
  the bed with the bottom wall.
\item The ratio of accelerations $\Gamma = (U_{0}\, \omega_{w} / g)$,
  which is proportional to the inverse of the ratio of frequencies,
  varies in a range $10^{2}$--$10^{4}$.
\item The ratio of the amplitude $a_{w}$ of the bottom wall vibrations
  to the mean free path varies in the range $5 \times 10^{-4}$--0.04, so
  that the wall does not induce any mean motion in the bed.
\end{enumerate}
In this parameter regime, the dynamics of the bed is coupled to that
of the bottom wall only through the energy transfer to the particles
(by random collisions with the wall), and the mean motion of the bed
is not correlated to the dynamics of the bottom wall. Therefore, the
periodicity of the bottom wall vibration does not induce a periodic
motion in the granular layer. This is confirmed by the observation that the
instabilities are observed even when the bottom wall is kept
stationary, but the wall velocity at collision is selected randomly
with a probability distribution identical to that for the vibrating
wall \citep{sun:thesis}.

The simulation cell is a box containing $N$ circular discs per unit
width vibrated at the bottom. The gravitational acceleration force $g$
acts vertically downwards.  There are two side walls separated by a
width $W$, and the top is open.  Inter-particle collisions are
inelastic, and are characterized by a constant coefficient of
restitution $e$.  For the sake of simplicity, the interaction of the
particles with the three walls is assumed to be elastic.

\begin{figure}[htb]
  \begin{center}
    \gpfig{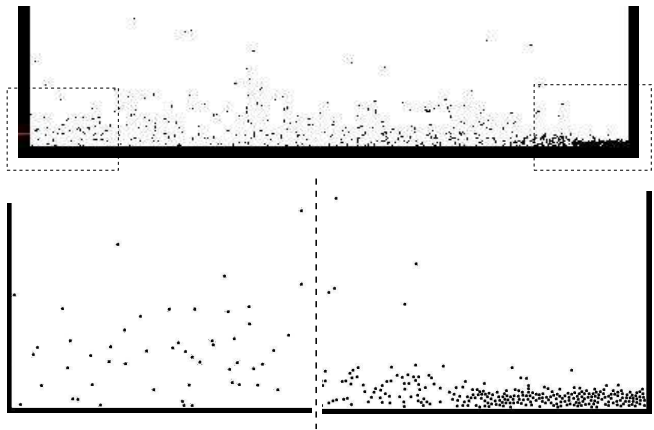}
    \caption{Exploded view of snapshots of regions near the left and
      the right wall.  A coexistence of the dilute and dense regions
      can be seen near the right wall. \param{1}{0.75}{0.077}. The
      horizontal width of the cell here is $W=624$. The variation in
      the scaled density fluctuation can be seen in
      \Figref{fig:nvar}.}
    \label{fig:clust}
  \end{center}
\end{figure}

\begin{figure}[tbp]
  \begin{center}
    \xlabel{$x$}
    \ylabel{$\tilde{N}_{x}$}
    \gpfig{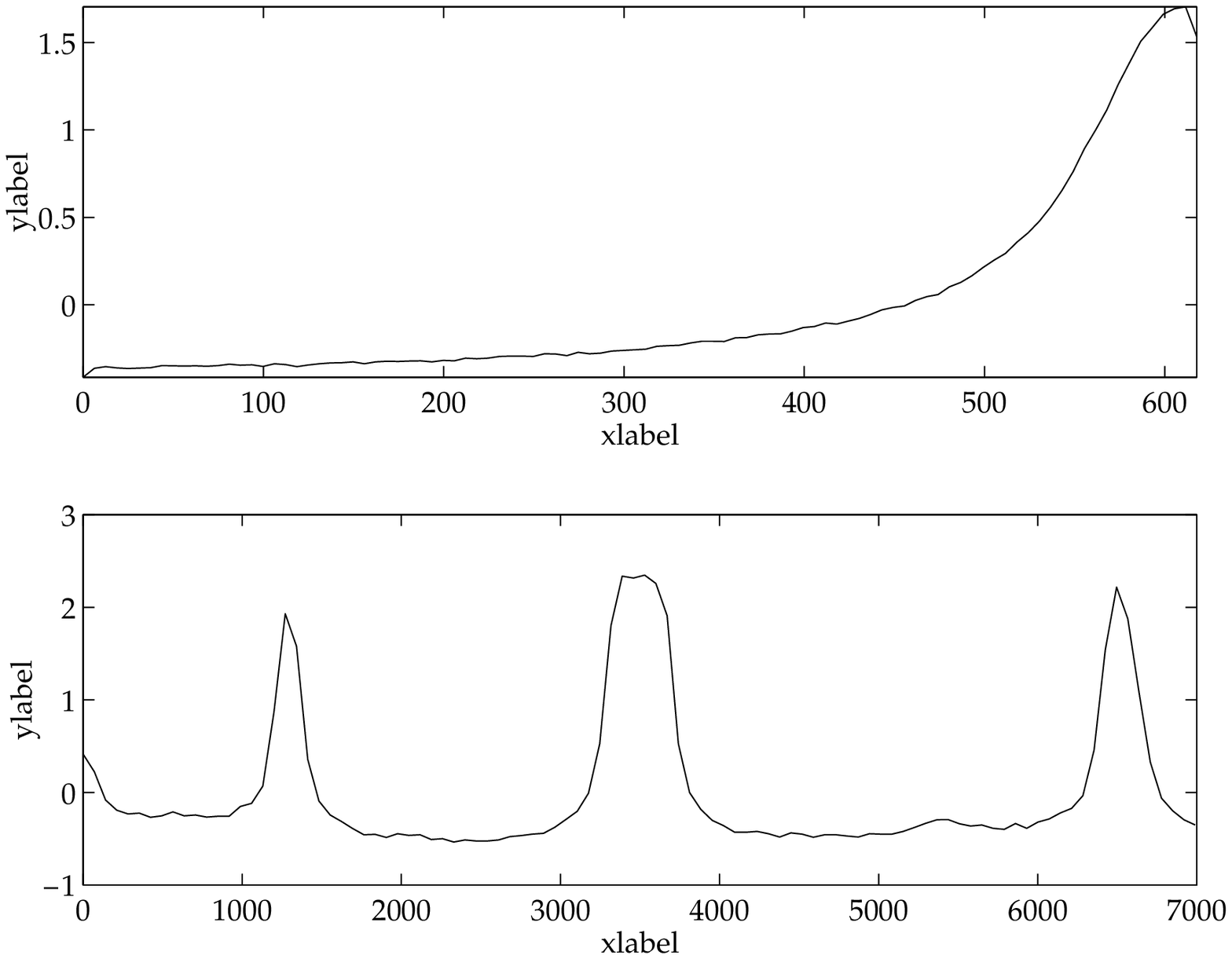}
    \caption{The steady state variation of the scaled density
      fluctuation $\tilde{N}(x)$ across the width of the cell is shown
      here for two different cell widths.  \param{1}{0.75}{0.077}.  A
      snapshot of the first plot is shown in \Figref{fig:nvar}. The
      dense regions coexist with dilute regions on both sides and are 
      not necessarily restricted near the walls of the container, 
      as shown in the second plot.} 
    \label{fig:nvar}
  \end{center}
\end{figure}

\begin{figure}[htbp]
  \begin{center}
    \psfrag{Dilute}{{\small Dilute}}
    \psfrag{Dense}{{\small Dense}}
    \xlabel{$r/\sigma$}
    \ylabel{$g(r)$}
    \gpfig{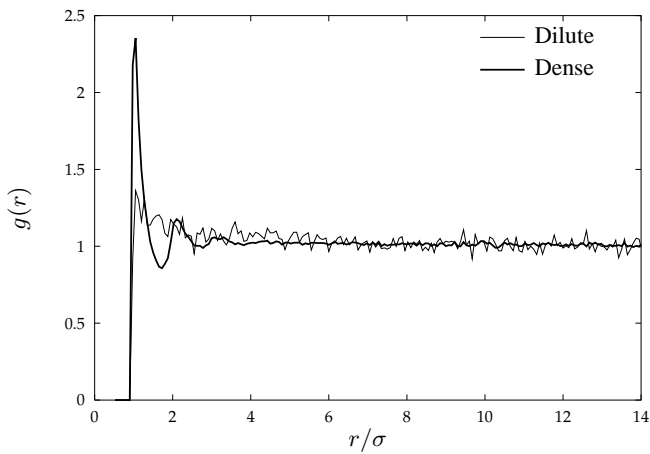}
    \caption{Pair distribution function in the dilute and dense
      phases. The static structure of the dilute phase is similar to a 
      gas and the dense phase is similar to a liquid.} 
    \label{fig:gofr-clus}
  \end{center}
\end{figure}

In the present study, we consider the effects varying $\ndp$ and
$\epsilon$; the other parameters are kept constant at the values
mentioned above.  The simulations show that at low values of the
parameter $\epsilon$, less than about $0.3$, there is a fluidized
state which is uniform in the horizontal direction. As $\epsilon$ is
increased above $0.3$ at constant $\ndp \leq 5$, there is a separation
of coexisting regions of high and low densities. (For $\ndp > 5$, the
system forms convection rolls, which will be reported elsewhere).
This can be seen in \Figref{fig:clust}, which shows the coexistence of
a dense and a dilute region.  After the formation of two coexisting
regions, there is no further variation in the average densities of the
two regions, though there is a variation in the position of the
interface between the two regions.  Changing the width of the cell
does not alter the length scale of the density fluctuation, but simply
alters the number of dense and dilute regions as in \Figref{fig:nvar},
which shows the variation of the scaled local density fluctuation
$\tilde{N}(x) = N_{x}/N -1$ for two different cell widths.  (It can
also be shown that there is no phase separation for smaller widths,
giving one possible reason why such instabilities were not observed
earlier.)
The static structure of the two phases can be represented
by the pair distribution function $g(r)$, which is shown in
\Figref{fig:gofr-clus}.  Here we see that the dilute region is like a
gas phase with uncorrelated particle positions and the dense phase is
like a liquid phase showing correlations up to a few particle
diameters.

\begin{figure}[htbp]
  \begin{center}
    \psfrag{x=0}{{\small Dilute}}
    \psfrag{x=850}{{\small Dense}}
    \xlabel{$u_{x}$}
    \ylabel{$f(u_{x})$}
    \gpfig{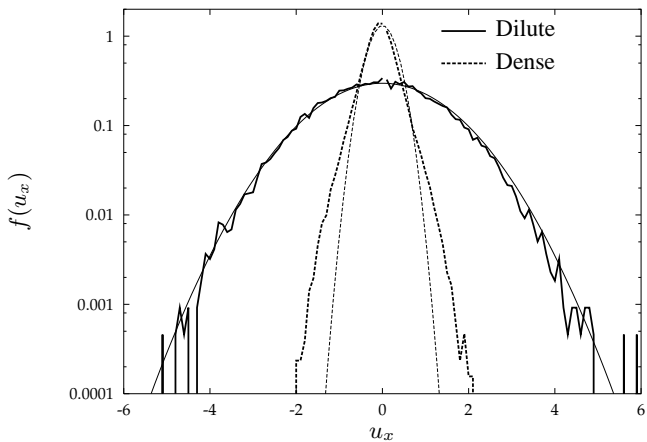}
    \caption{The distribution function of horizontal velocities in the
      dilute and the dense regions is plotted here.  A Gaussian curve
      corresponding to the same local temperature $T$ is also plotted
      along for comparison.  It is seen that for the dilute region
      distribution function is very close to the Gaussian curve,
      whereas for the dense region the distribution function shows a
      non-Gaussian behavior.}
    \label{fig:nongauss}
  \end{center}
\end{figure}

Another salient feature of this coexistence is the form of the velocity
distribution function in the dense phase.  \Figref{fig:nongauss} shows the
velocity distribution functions in the dilute and the dense regions.
A Gaussian curve with the same temperature is also plotted along each for 
reference.  It is seen that in the dilute region,
the distribution function is very close to the Gaussian distribution,
whereas in the dense region the distribution function shows
non-Gaussian tails.  Such non-Gaussian distributions have been found
in many other granular systems where other kinds of clustering
phenomena have been found \citep{\nongauss}.

It is to be noted that the two phase state is not observed if a
constant temperature condition is used at the bottom wall for the same
parameter values as those for which phase separation is observed with
a vibrating wall. The physical reason for this is as follows. When a
constant temperature condition is enforced at the bottom wall, the
mean kinetic energy of the particles after collision with the wall is
independent of the kinetic energy before collision. This boundary
condition tends to damp out temperature variations in the horizontal
direction, since the post-collisional mean kinetic energy of the
particles is independent of horizontal position. When a real vibrating
wall is used, however, the collision of a particle with the wall only
augments the kinetic energy of a particle, and so horizontal
temperature variations are not damped out at the wall. This leads to
the coexistence of regions with two different temperatures for a
system driven by a vibrated surface for conditions under which a
system with a constant temperature surface does not show this
coexistence.

\begin{figure}[tbp]
  \begin{center}
   \psfrag{x2label}[][]{$x$} \psfrag{y2label}[b][t]{Time}
   \psfrag{x1label}[][]{$\omega$} \psfrag{y1label}[b][t]{$x$}
   \psfrag{zlabel}[][]{FFT}
   \gpfig{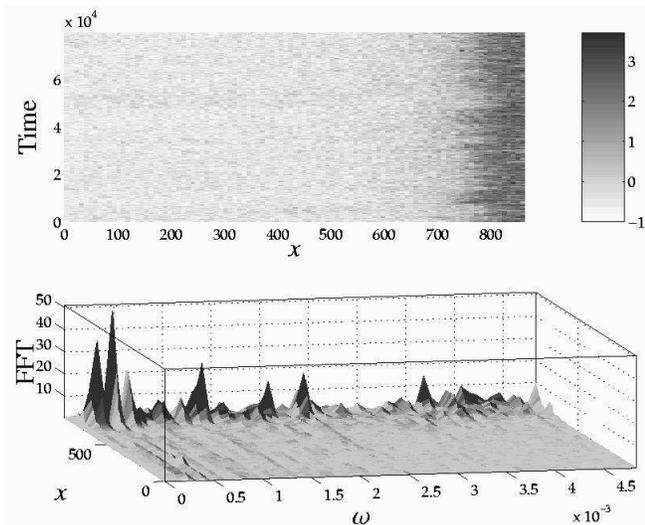}
    \caption{Dynamics of the interface. A close inspection of the
      local scaled density fluctuations $\tilde{N}(x)$ in top plot
      shows that the cluster interface is not static but varying with
      time. The second plot shows the Fourier transform (FFT) of the
      density fluctuation in the time domain at different horizontal
      positions. A peak is found at $x=770$ near the interface region.
      The characteristic radian frequency is around $4\times 10^{-4}$,
      which corresponds to a time period of $1.6\times 10^{4}$ roughly
      of the order of interface fluctuation in the top plot.}
    \label{fig:dyna-clus}
  \end{center}
\end{figure}

The simulations also indicate that the horizontal position of the
interface between the high and low density regions fluctuates in time,
as shown in \Figref{fig:dyna-clus}. To study the dynamics of the
interface, we have plotted the Fourier transform of the density
fluctuation at the interface.  In the second plot of
\Figref{fig:dyna-clus}, we see that the dominant peak occurs only in
the interface region and is clearly characterized by a single
frequency, the corresponding time period ($1.6\times 10^{4}$) matches
the order of fluctuations seen in the time domain.

It is natural to attempt to provide an explanation similar to
\cite{eggers99} for the coexistence of two densities in a vibrated bed
separated into two compartments with a hole in the wall. However, it
is shown here that such an explanation is not valid for the present
case.  In the spirit of \cite{eggers99}, let us consider the
possibility of a distribution of the particles into two
``compartments'' with two different densities (number of particles per
unit width) $N_{1} < N_{2}$.  Coexistence requires that the horizontal
fluxes of particles $F_{1}$ and $F_{2}$ between these two compartments
be equal.  The width $W_{1}$ of the first compartment can be
determined from the constraint on the total number of particles. The
constraints for a steady state distribution are
\begin{eqnarray}
  \label{eq:hflux-constraints}
  &F_{1}  =  F_{2} \\
  &N_{1}\,W_{1} + N_{2}\,(W-W_{1})  =  N\,W
\end{eqnarray}
The above condition requires that the total horizontal flux $F(N)$ be
a non-monotonous function of $N$.  This can be realized when the
exchange between the two compartments takes place only at a particular
height, as shown in \cite{eggers99}.  When there is no wall separating
the two compartments, the total flux of particles across any vertical
surface is $F(N) = N\, \textfrac{\sqrt{T}}{2\pi}$.  Since $T \sim 1/N$
from \Eqref{eq:T0}, we have $F\sim\sqrt{N}$ which is a monotonic
function of $N$.  Therefore, a low density theory cannot explain the
existence of a steady distribution.  The flux at high density has to
be obtained using the density profiles computed numerically by
including the virial equation of state for the pressure---density
relationship instead of the ideal gas equation of state
\citep{sunkum99:scal}.  It turns out that the variation of the fluxes
for higher densities increases faster than $\sqrt{N}$.  We have also
computed the horizontal fluxes from simulation, by calculating the
rate of particle collisions with the side walls for a system with
different $N$ and keeping all other parameters at a constant. For this
simulation the width of the system is taken to be smaller than the
length scale of separation so that the bed does not develop any
inhomogeneity in the horizontal direction.  The simulation results
also indicate that the flux increases faster than $\sqrt{N}$ in
agreement with the theoretical predictions.  We have also verified
that the simulations, in the case of the phase separation, do not
violate the net zero flux condition across different vertical
surfaces. This implies that the coexistence of two phases cannot be
understood using simple steady state flux balance models such as the
one presented above, in contrast to the clusters (or phase separation)
of \cite{eggers99}, which could be explained using the dilute bed
kinetic theory of \cite{kum98:vibscal}.  However, it can be shown
\citep{sun:thesis} that the variation of the density and the
temperature in the homogeneous state just before the occurrence of this
instability can be adequately represented by the kinetic theory of
\cite{kum98:vib}, where the effects of the periodicity of wall
vibration are neglected. This indicates that the stability of the 
system can be understood in terms of simple equations.

To conclude, a spontaneous symmetry breaking phenomenon is observed
in a vibrated bed in a regime where it has not been observed before,
i.e., in the dilute limit where the area fraction is
small, the amplitude of the vibrating surface is small compared to
the mean free path and the frequency is large compared to the frequency
of mean motion of the bed.   This appears similar
to that reported by \cite{eggers99}, but a detailed analysis indicated
that the present phenomenon cannot be explained the simple model used
there.  This
requires a more complicated model where the fluctuating nature of the
interface and possibly the non-Gaussian nature of the distribution
function in the dense phase is incorporated.  A stability analysis of
this state could have implications in understanding instabilities in
driven granular materials and in pattern formation in vibrated beds.


\begin{thebibliography}{10}

\bibitem{jae-nag96}
H.~M. Jaeger and S.~R. Nagel, Rev. Mod. Phys. {\bf 68},  1259  (1996).

\bibitem{waves}
H.~K. Pak and R.~P. Behringer, Phys. Rev. Lett. {\bf 71},  1832  (1993);
E. Cl\'ement, L. Vanel, J. Rajchenbach, and J. Duran, Phys. Rev. E {\bf 53},
  2972  (1996).


\bibitem{umbanetal96}
P.~B. Umbanhowar, F. Melo, and H.~L. Swinney, Nature {\bf 382},  793  (1996).

\bibitem{meloetal95}
F. Melo, P.~B. Umbanhowar, and H.~L. Swinney, Phys. Rev. Lett. {\bf 75},  3838
  (1995).

\bibitem{venkat-ott98}
S.~C. Venkataramani and E. Ott, Phys. Rev. Lett. {\bf 80},  3495  (1998).

\bibitem{eggers99}
J. Eggers, Phys. Rev. Lett. {\bf 83},  5322  (1999).

\bibitem{kum98:vibscal}
V. Kumaran, Phys. Rev. E {\bf 57},  5660  (1998).

\bibitem{buevich-ryzhkov80}
Y.~A. Buevich and A.~F. Ryzhkov, Jl. of Engng. Phys. and Thermal Phys. {\bf
  626},  1162 (English)  (1980), translated from Russian.

\bibitem{phases-sim}
D.~R.~M. Williams, Physica A {\bf 233},  718  (1996);
X. Nie, E. Ben-Naim, and S.~Y. Chen, Europhys. Lett. {\bf 51},  679  (2000).

\bibitem{phases-expt}
J.~S. Olafsen and J.~S. Urbach, Phys. Rev. Lett. {\bf 81},  4369  (1998);
W. Losert, D.~G.~W. Cooper, and J.~P. Gollub, Phys. Rev. E {\bf 59},  5855
  (1999).

\bibitem{sun:thesis}
P. Sunthar, Ph.D. thesis, Indian Institute of Science, Bangalore, India, 2001,
  \granuhome.

\bibitem{nongauss}
A. Puglisi {\it et~al.}, Phys. Rev. Lett. {\bf 81},  3848  (1998);
J.~S. Olafsen and J.~S. Urbach, Phys. Rev. E {\bf 60},  R2468  (1999);
F. Rouyer and N. Menon, Phys. Rev. Lett. {\bf 85},  3676  (2000).

\bibitem{sunkum99:scal}
P. Sunthar and V. Kumaran, Phys. Rev. E {\bf 60},  1951  (1999).

\bibitem{kum98:vib}
V. Kumaran, J. Fluid Mech. {\bf 364},  163  (1998).

\end{thebibliography}

\end{document}